\documentclass[aps,pra, twocolumn,superscriptaddress,floatfix]{revtex4-1}
\usepackage[english]{babel}
\usepackage{amsmath,amsfonts,amssymb}
\usepackage{graphicx}
\usepackage{epstopdf}
\usepackage{natbib}

\let\originalleft\left
\let\originalright\right
\renewcommand{\left}{\mathopen{}\mathclose\bgroup\originalleft}
\renewcommand{\right}{\aftergroup\egroup\originalright}

\newcommand{\<}{\langle}
\renewcommand{\>}{\rangle}

\renewcommand{\<}{\left\langle}
\renewcommand{\>}{\right\rangle}

\begin{document}
\title{A Statistical Benchmark for BosonSampling}
\author{Mattia Walschaers} 
\email{mattia@itf.fys.kuleuven.be}
\affiliation{Physikalisches Institut, Albert-Ludwigs-Universitat Freiburg, Hermann-Herder-Str. 3, D-79104 Freiburg, Germany}
\affiliation{Instituut voor Theoretische Fysica, University of Leuven, Celestijnenlaan 200D, B-3001 Heverlee, Belgium}
\author{Jack Kuipers}
\affiliation{Institut f\"ur Theoretische Physik, Universit\"at Regensburg, D-93040 Regensburg, Germany}
\affiliation{D-BSSE, ETH Z\"urich, Mattenstrasse 26, 4058 Basel, Switzerland}
\author{Juan-Diego Urbina}
\affiliation{Institut f\"ur Theoretische Physik, Universit\"at Regensburg, D-93040 Regensburg, Germany}
\author{Klaus Mayer}
\affiliation{Physikalisches Institut, Albert-Ludwigs-Universitat Freiburg, Hermann-Herder-Str. 3, D-79104 Freiburg, Germany}
\author{Malte Christopher Tichy}
\affiliation{Department of Physics and Astronomy, Aarhus University, Ny Munkegade 120, DK-8000 Aarhus, Denmark}
\author{Klaus Richter}
\affiliation{Institut f\"ur Theoretische Physik, Universit\"at Regensburg, D-93040 Regensburg, Germany}
\author{Andreas Buchleitner}
\affiliation{Physikalisches Institut, Albert-Ludwigs-Universitat Freiburg, Hermann-Herder-Str. 3, D-79104 Freiburg, Germany}
\email{andreas.buchleitner@physik.uni-freiburg.de.}

\date{\today}

\begin{abstract}
Computing the state of a quantum mechanical many-body system composed of indistinguishable particles distributed over a multitude of modes
is one of the paradigmatic test cases of computational complexity theory: Beyond well-understood quantum statistical effects, the coherent superposition of
{\em many-particle} amplitudes rapidly overburdens classical computing devices - essentially by creating extremely complicated interference patterns, which
also challenge experimental resolution. With the advent of controlled many-particle interference experiments, optical set-ups that can efficiently probe many-boson
wave functions - baptised BosonSamplers - have therefore been proposed as efficient quantum simulators which outperform any classical computing device, and thereby challenge
the extended Church-Turing thesis, one of the fundamental dogmas of computer science. However, as in all experimental quantum simulations of truly complex systems,
there remains one crucial problem: How to certify that a given experimental measurement record is an unambiguous result of sampling bosons rather than fermions or distinguishable
particles, or of uncontrolled noise? In this contribution, we describe a statistical signature of many-body quantum interference, which can be used as an experimental (and classically computable) benchmark for BosonSampling.
\end{abstract}

\maketitle

\section{Introduction} As the world waits for the first universal and fully operational quantum computer \cite{nielsen_quantum_2010}, quantum information scientists are eager to already show the power of quantum physics to perform computational tasks which are out of reach for a classical computer. Rather than designing devices that can perform a wide range of calculations, machines which are specialised in specific tasks have joined the scope \cite{aaronson_computational_2013,PhysRevX.4.021041}. Here, we focus on one type of such devices, the BosonSamplers, which may hold the key to falsifying the extended Church-Turing thesis \cite{nielsen_quantum_2010,deutsch_quantum_1985,MooreMertens}. This conjecture, rooted in the early days of computer science, states that any efficient calculation performed by a physical device can also be performed in polynomial time on a classical computer. It is now proposed that all that is necessary to falsify this foundational dogma of computer science is a set of $m$ photonic input modes, which are connected to $m$ output modes by a random photonic circuit \cite{aaronson_computational_2013}. This immediately indicates why BosonSampling attracts such attention, as these systems are experimentally in reach \cite{broome_photonic_2013,crespi_integrated_2013,ralph_quantum_2013,spagnolo_experimental_2014,spring_boson_2013,tillmann_experimental_2013}. 

\begin{figure}[t!]
  	\includegraphics[width=0.49\textwidth]{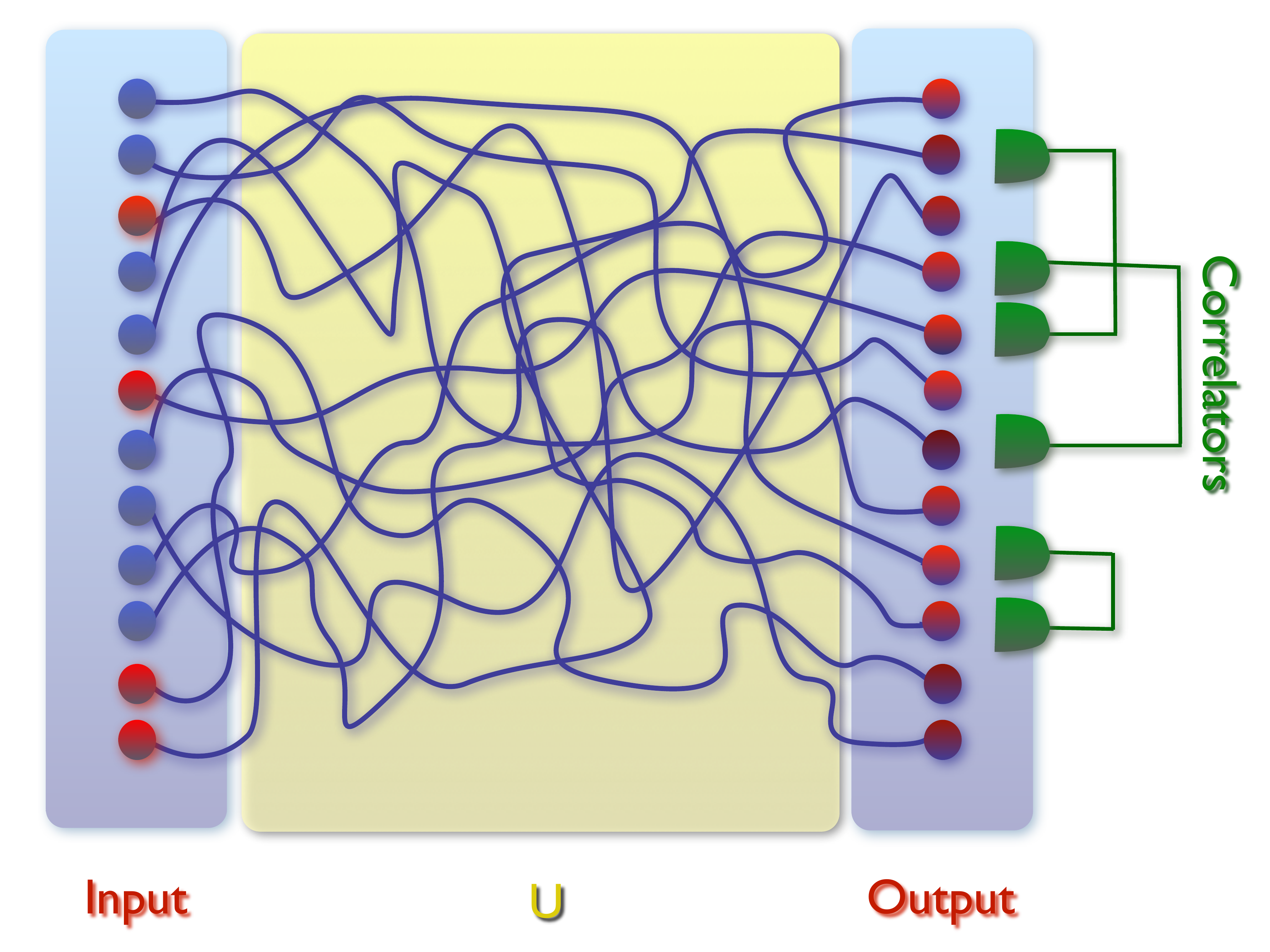}
  \caption{Sketch of the system under consideration. From a set of input modes (here $m=12$), four particles are injected (depicted in red). The particles traverse the system via the depicted channels. At the crossings, different paths are interconnected such that particles can travel on in each direction, thus leading to single-particle {\em and} many-body interference. The yellow zone of the setup is described by a unitary matrix $U$, which connects the input modes to the output modes. The output signal is probabilistic in nature (hence different intensities of red), with its statistics governed by the many-body quantum state. The key object of this work, the C-dataset (see main text), is obtained by calculating the two-point correlations between different modes as depicted in green.}
   \label{fig:Sketch}
\end{figure}

Since we are dealing with bosons, indistinguishable particles, interesting physics arises when multiple particles are simultaneously injected into the system \cite{hong_measurement_1987,tichy_zero-transmission_2010,mayer_counting_2011,tichy_many-particle_2012,ra_nonmonotonic_2013}. A schematic overview, indicating the essential ingredients of a sampling device such as considered here, is provided in Figure \ref{fig:Sketch}. BosonSampling essentially consists in quantifying the probability of measuring an occupation vector $\vec{y} = (y_1, \dots, y_m)$ for the output modes, given that the initial mode occupation was $\vec{x} = (x_1, \dots x_m)$, where $x_i$ and $y_i$ are the number of photons found in the $i$th input and output mode, respectively. In the case with at most one photon per input mode, this probability is given by $p_{\vec{x} \rightarrow \vec{y}} = \frac{\lvert{{\rm perm}~U_{\vec{x}, \vec{y}}\rvert}^2}{\prod^m_{i=1} y_i!}$, where $U_{\vec{x}, \vec{y}}$ is the $n \times n$ matrix, constructed from $U$, which describes how the $n$ occupied input modes are connected to the selected output modes \cite{tichy_zero-transmission_2010}, and ``${\rm perm}$'' denotes the {\em permanent} \cite{minc_permanents_1978}. The latter lies at the heart of the contemporary interest in the BosonSampling problem: In order to evaluate the many-body wave function, and thus the sampling statistics, we must calculate permanents. A permanent, however, is ``hard'' to compute, as there is no algorithm that can do so in polynomial time \cite{aaronson_computational_2013, minc_permanents_1978,troyansky_quantum_1996}. Thus, the BosonSampler is a physical system that efficiently samples bosons according to the bosonic many-body wave function, even though this many-body state {\em cannot} be calculated by a classical computer in polynomial time. Its realisation would in this perspective invalidate the extended Church-Turing thesis. This strength of BosonSampling apparently also implies a profound weakness: As it is impossible to  compute the many-body wave function, one cannot certify that a given experimental output unambiguously stems from sampling bosons.  

Rather than joining in the mathematical debate about the similarity of BosonSampling to sampling over other distributions \cite{aaronson_bosonsampling_2013,gogolin_Boson-Sampling_2013} or setting a benchmark based on an analytically treatable instance outside the computationally hard generic case \cite{tichy_stringent_2014}, we here tackle the problem from a  {\em physics perspective} inspired by statistical mechanics. Borrowing from the vast set of techniques applied in this field, we study {\em transport processes} in scattering systems ({\em e.g.} a photonic network), given by a unitary matrix $U$. We show that, by combining quantum optics with methods found in statistical physics and random matrix theory, it is indeed possible to identify signatures of genuine bosonic interference with manageable overhead.

\section{Statistical Signatures of Interference} It must be stressed that the resulting many-body wave function of a scattering process, such as manifested in a BosonSampler, is determined by several factors. One con divide them into those which are of a statistical origin and those which are dynamical in nature. The former concern all effects related  to the quantum statistics ({\em e.g.}~the Pauli exclusion principle), whereas the latter involve all sorts of interference effects. One can divide these interferences into the well-known single-particle interference \cite{Akkermans2007}, encoded within the matrix $U$, which arises due to the wave-like nature of quantum transport \cite{Bloch:1929aa,PhysRevA.48.1687,scholak_spectral_2014}, and the far more difficult many-body interference \cite{tichy_many-particle_2012, ra_nonmonotonic_2013, PhysRevLett.112.140403}. Here, we shall study the sampling of various particle types -- bosons, fermions, distinguishable particles and ``simulated bosons''-- and explain how to efficiently distinguish between them. Distinguishable particles are the simplest of the considered species, as their signals are governed only by single-particle interference. Transport processes with bosons or fermions, in contrast, exhibit the entire range of statistical and interference effects. As indistinguishable particles, they obey quantum statistics (as dictated by the relevant algebra), which for two particles in two modes either leads to {\em bunching} (bosons) \cite{hong_measurement_1987} or {\em anti-bunching} (fermions) \cite{Jeltes:2007aa}. Similar effects have also been observed in larger setups, with more particles, and have hence been proposed as hallmarks for many-boson \cite{carolan_experimental_2014} or many-fermion \cite{Rom:2006aa} behaviour.  However, for larger setups one expects a much richer phenomenology due to many-body interference \cite{tichy_many-particle_2012}. For BosonSampling, bunching behaviour as such is not a sufficient tool for certification, as it can easily be achieved by the so-called mean-field sampler  \cite{tichy_stringent_2014} which was inspired by semiclassical models \cite{chuchem_quantum_2010}. These devices are designed to replicate the bunching behaviour by approximating the many-body output state by a macroscopically populated single-particle state with random phases added. Averaging over the random phases mimics boson-like bunching. However, all relative phases are destroyed in the averaging procedure and therefore all many-body interference effects are vanquished. This type of sampling is easily simulated by Monte Carlo methods -- hence we refer to such sampled particles as ``simulated bosons''-- stressing the importance of setting this type of sampling apart from the actual BosonSampling. This, however, can be done, since we here introduce a method to successfully differentiate the transport processes of all these different particle types.

\section{Random Matrix Methods } The scattering matrix $U$ that describes the photonic circuit in the BosonSampling setup is randomly sampled from the {\em Haar measure} \cite{mehta_random_2004,mezzadri_how_2006,zyczkowski_random_1994}. Therefore it is only natural to treat this problem in a framework of statistics and random matrix theory (RMT) \cite{samuel80,mello90,brouwer_diagrammatic_1996,mehta_random_2004,berkolaiko_combinatorial_2013}. Often the lack of grasp on the statistical distribution of the full many-body wave function is put forth as the core of the certification problem. Exhaustive statistical characterisation of the many-body state would require the full  distribution of permanents over the set of unitary matrices. To date, only the first moment of this distribution is known \cite{urbina_multiparticle_2014}  and it is not enough to provide certification, while sufficiently precise higher order  moments are out of reach \cite{aaronson_computational_2013} .  The reason is that, in terms of the quantum state, permanents depend on high-order correlation functions. We, on the other hand, want to emphasise the gigantic amount of information about the many-body state which {\it is} within reach in the form of distributions of low-order correlation functions.

In probability theory, the knowledge of all possible correlation functions implies full knowledge of the (joint) probability distribution itself. Therefore, correlation functions play a central role in many probabilistic theories, from RMT \cite{mehta_random_2004} to quantum statistical mechanics \cite{bratteli_operator_1997}. In practical applications of RMT, such as occur in quantum chaos, the full joint probability distribution of many eigenvalues of a large random matrix is out of reach; however, a study of the statistics related to two-point correlations is often sufficient to certify the RMT ensemble. Similarly, here, we do not know {\em all} correlation functions of the many-body quantum state. There is, however, a large set of correlation functions which {\em are accessible} (both theoretically and experimentally) \cite{Peruzzo:2010wt,mayer_counting_2011,urbina_multiparticle_2014}. Hence, the only relevant question is whether this offers a sufficient amount of information on the many-body states for the various particle types to be distinguished. We show that the answer to the question is affirmative.

\section{Statistical Benchmarking}  We propose the mode correlator \cite{mayer_counting_2011}, $C_{ij}=\<\hat{n}_i\hat{n}_j\> -\<\hat{n}_i\>\<\hat{n}_j\> $ (with $\hat{n}_i = a_i^*a_i$ the bosonic number operator), which quantifies how the number of outgoing photons in modes $i$ and $j$ are correlated (as indicated in Figure \ref{fig:Sketch}), as a suitable quantity to differentiate particle types. The particle type is encrypted in the many-body output quantum state $\left\lvert\phi_{\rm out} \>$, which shows up in $C_{ij}$ via the expectation value $\<.\> = \<\phi_{\rm out}\right\rvert.\left\lvert\phi_{\rm out} \>$. A single such correlator cannot (typically) offer much insight in the full wave function, rather some characteristic behaviour will become apparent once a sufficiently large set of such correlators is considered. This type of dataset, which we further refer to as the {\em C-dataset}, is easily obtained in the experimental setup: One must consider all possible choices for output modes, $i,j \in\{1, \dots m\}$ such that $i < j$, and for each choice compute the correlation $C_{ij}$ between the number of particles sampled in the two modes. Now, for a {\em single} choice of input modes and hence a {\em single} $n \times m$ unitary matrix $U_{\rm sub}$, the submatrix of $U$ that describes how the input modes are coupled to all possible output modes, we obtain a set of data on which we can do statistics.

Given $U_{\rm sub}$, it is possible to exactly numerically calculate the C-dataset for each particle type \cite{mayer_counting_2011} (see Methods). Although we can explore this numerically via histograms, moments and other statistical properties, it is far from straightforward to get an analytical grasp of its statistics. Ideally, we would like to predict the exact shape of the distribution, given the number of input modes $m$ and the number of incoming particles $n$, but this appears to be an unrealistic goal. Nevertheless, after longwinded RMT calculations, we obtained analytical predictions for the first three moments of the set of possible outcomes when varying $U_{\rm sub}$ for fixed $i$ and $j$ (rather than fixing $U_{\rm sub}$ and varying $i$ and $j$). Although the distributions are mathematically not exactly equivalent, we find good agreement with numerics (for a more elaborate discussion, see Methods).

\begin{figure}
  	\includegraphics[width=0.49\textwidth]{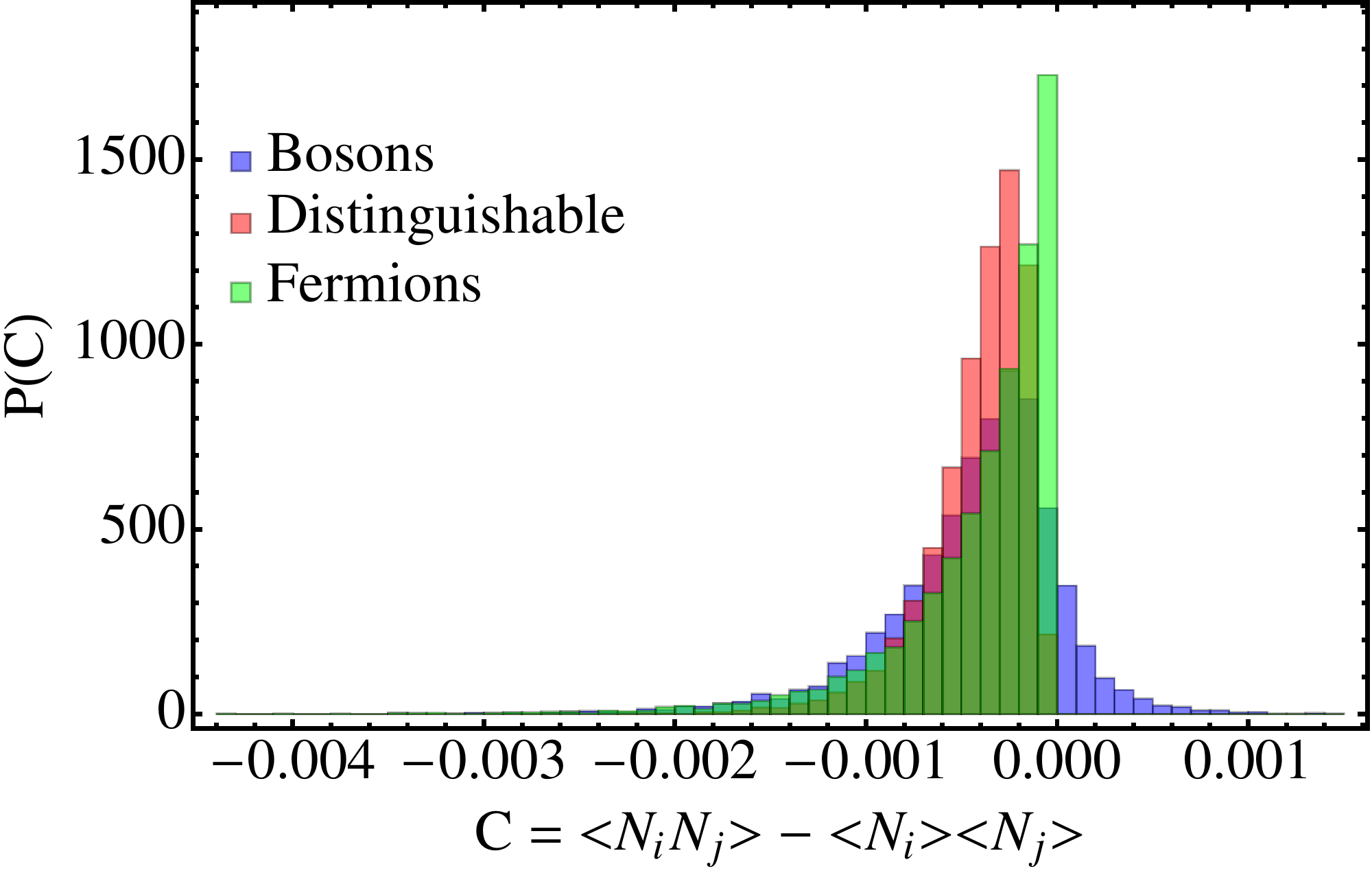}
	\includegraphics[width=0.49\textwidth]{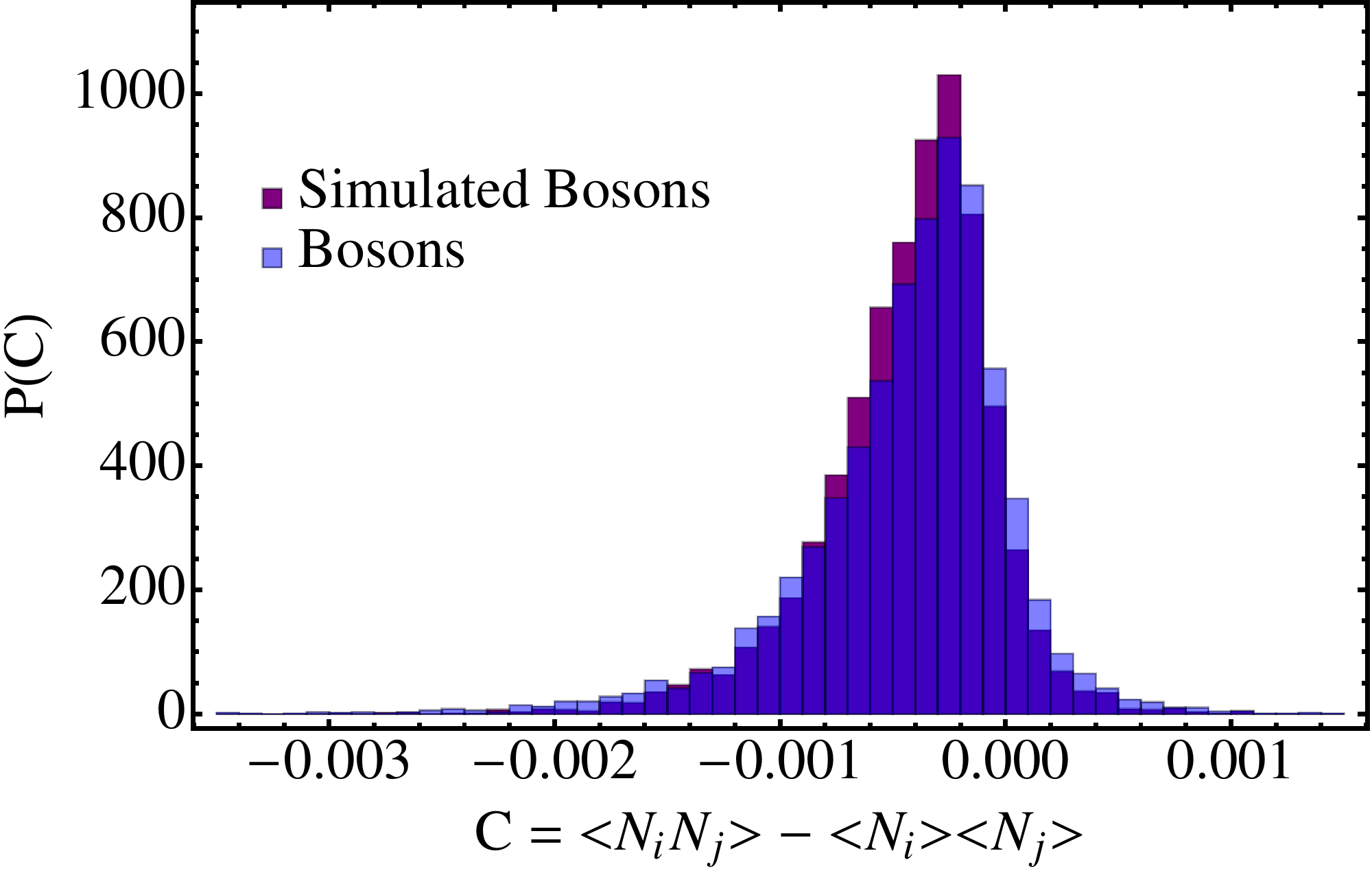}
  \caption{Normalised histograms of the correlator data, obtained by computing $C_{ij}=\<\hat{n}_i\hat{n}_j\> -\<\hat{n}_i\>\<\hat{n}_j\> $ for all possible mode combinations, for a system with six particles and 120 modes. Top panel, the histogram for bosonic correlators is compared to data obtained with fermions or distinguishable particles inserted instead of bosons. At the bottom, the histograms for bosons is compared to the result for simulated bosons, see main text. All histograms are obtained from one single circuit, using the same input modes, thus implying the same $U_{\rm sub}$. }
   \label{fig:Hists}
\end{figure}

For a generated C-dataset for the different particle types, considering one fixed $U_{\rm sub}$ for $120$ output modes and six particles, the top panel in Figure \ref{fig:Hists} clearly shows a qualitative difference in the histograms for different particle types. In contrast, the bottom panel of Figure \ref{fig:Hists} indicates that the histograms of the true bosons and their simulated counterparts bear a strong resemblance. Obviously, a quantitative understanding is essential to really distinguish bosons from the other particle species. The second and third moment of the obtained correlator dataset can exactly provide us with such an insight. Obtaining these quantities involves averaging products of components of unitary matrices, for which straightforward (but tedious) combinatorics are used \cite{berkolaiko_combinatorial_2013,brouwer_diagrammatic_1996}.

\section{Differentiating Particle Species} 
\begin{figure}
\includegraphics[width=0.49\textwidth]{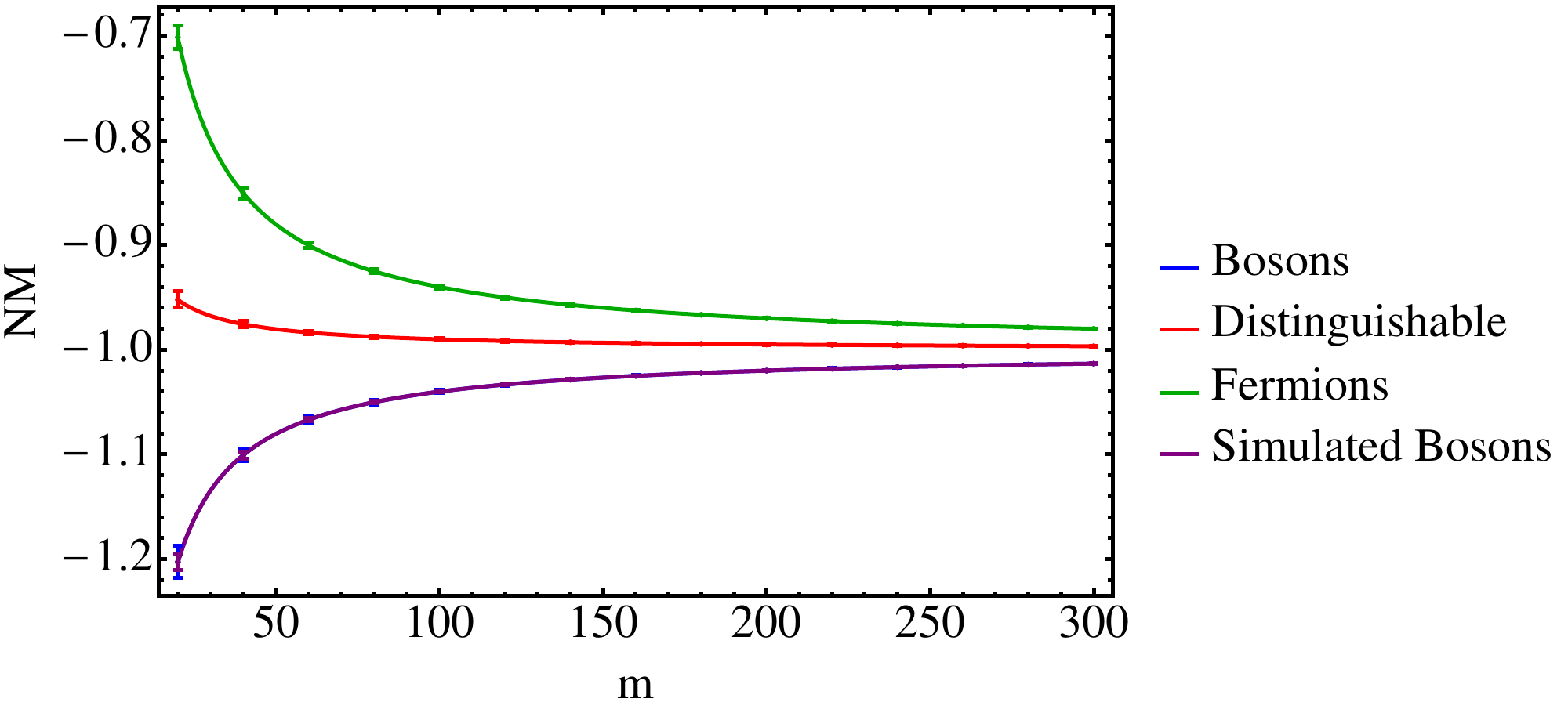}
\includegraphics[width=0.49\textwidth]{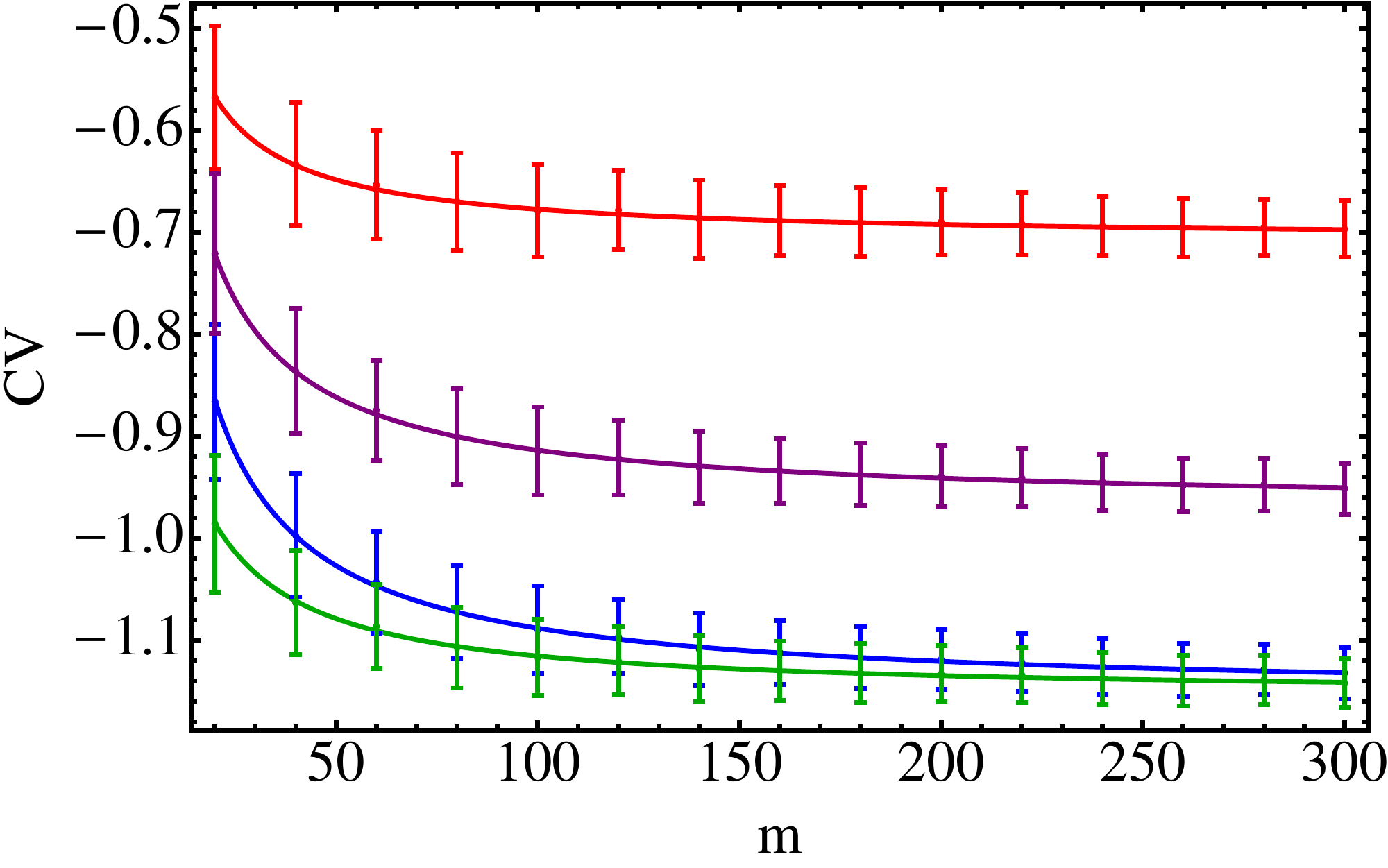}
\includegraphics[width=0.49\textwidth]{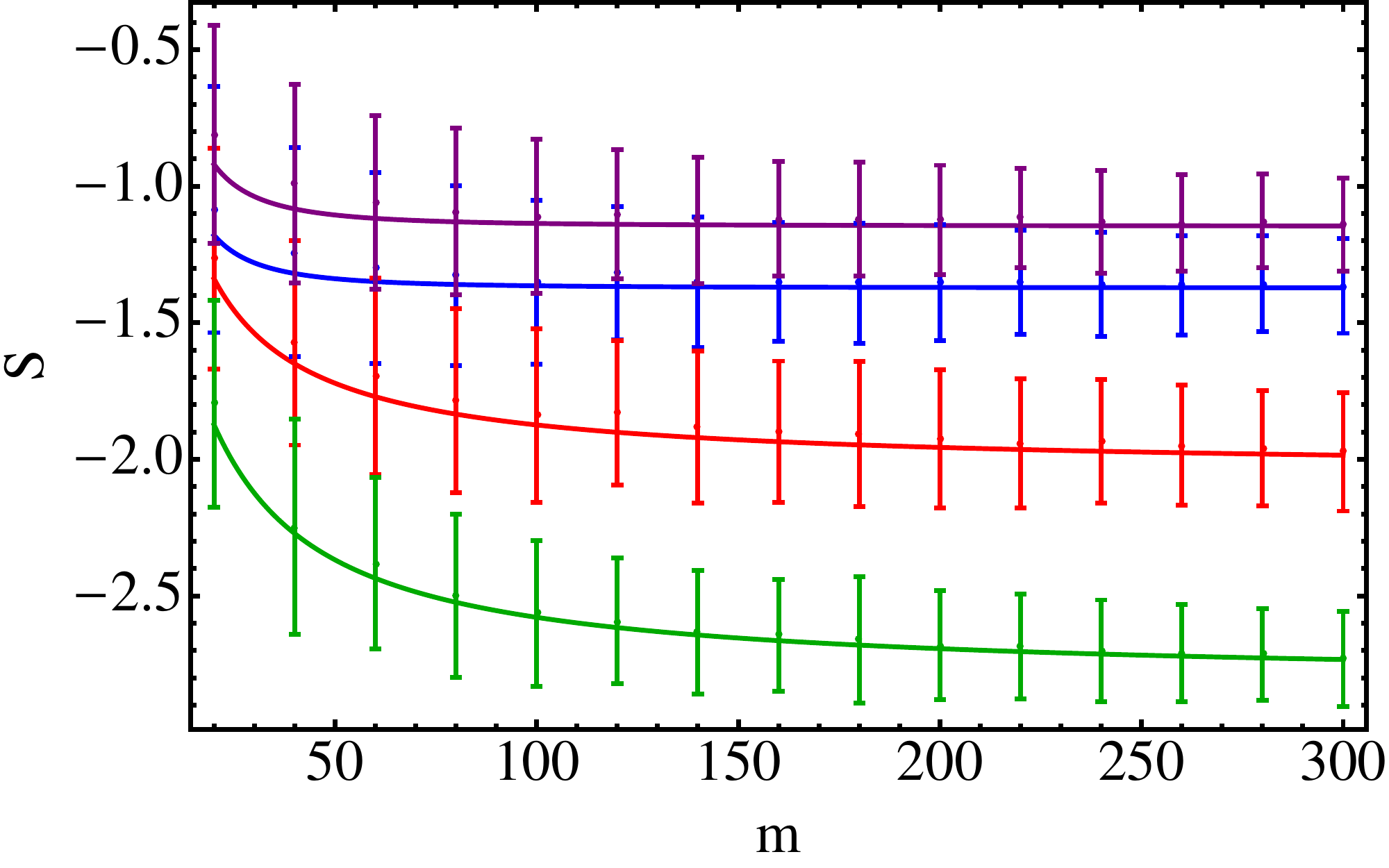}
  \caption{The theoretical RMT predictions (solid lines) for the normalised mean $NM$ (top), the coefficient of variation $CV$ (middle) and skewness $S$ (bottom) of the C-dataset are compared to the numerical $NM$, $CV$ and $S$ values of sampled C-datasets for six particles, as a function of the number of modes $m$. For mode numbers $m=20, 40, \dots, 300$, we sampled $500$ matrices $U_{\rm sub}$, for each of which the normalised mean, the coefficient of variation and skewness of the C-dataset were calculated. For each mode number, the average normalised mean, coefficient of variation and  skewness are indicated by a dot. Additionally, the standard deviations of the obtained $NM$, $CV$ and $S$ results are shown by error bars around these dots.}
   \label{fig:TheovsNum}
\end{figure}

To acquire the clearest distinction between different particle types, we propose the  {\em normalised mean} ($NM$) -- the first moment divided by $n/m^2$, the {\em coefficient of variation} ($CV$) -- the standard deviation divided by the mean -- and {\em skewness} ($S$) \cite{macgillivray_skewness_1986} of the C-dataset as benchmarks. For these quantities, we have obtained an analytical RMT prediction in terms of mode and particle number, but as these expressions are rather longwinded, we present them in the Appedix. Since the dataset is generated for a single $U_{\rm sub}$, as explained before, we do expect slight deviations from these RMT results. In Figure \ref{fig:TheovsNum}, we show the theoretical predictions (solid lines) for $NM$, $CV$ and $S$, for a sampler in which six particles were injected, as a function of the number of modes. In order to quantify the deviations from the the RMT prediction, we sampled, for various numbers of modes, 500 different $U_{\rm sub}$ matrices, calculated the respective C-dataset and its moments, and indicated the resulting average coefficient of variation and skewness by a point. The error bars indicate the standard deviation from this mean value and thus quantify the typical spread of possible outcomes. We show here that for these parameters, even $NM$ is fit to effectively differentiate bosons, fermions and distinguishable particles, the curves for simulated and true bosons, however, collapse. $CV$, on the other hand, is a trustworthy quantity to distinguish true bosons from distinguishable particles and even from simulated bosons. The skewness $S$ completes the certification that the particles under consideration are actually bosons.

\begin{figure}
	\includegraphics[width=0.49\textwidth]{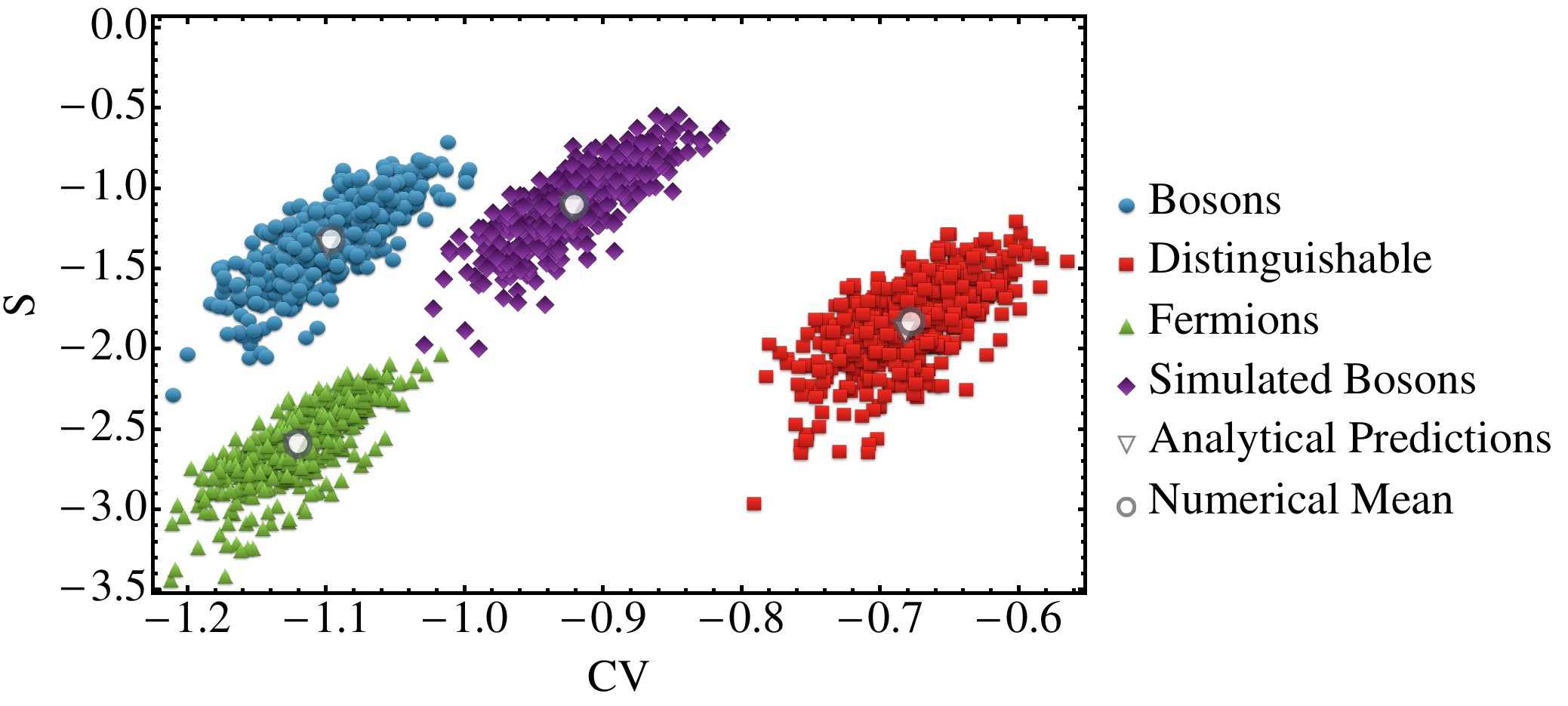}
  \caption{For six particles, in 120 modes, the coefficient of variation and skewness were calculated for C-datasets of $500$ sampled $U_{\rm sub}$ matrices, The points labeled ``bosons'', ``distinguishable'', ``fermions'' and ``simulated bosons'', each connect to one  C-dataset of a sampled $U_{\rm sub}$, the points' position on the plot indicates the calculated coefficient of variation $CV$ and skewness $S$. The black and white triangles mark the RMT predictions for these quantities. Finally, the black and white circles indicate the mean value of each cluster of all the points scattered of each particle type. These means coincide with the RMT predictions, thus the black and white triangles are hidden underneath the black and white circles.}
   \label{fig:VarSkew}
\end{figure}

We must emphasise that the curves of Figure \ref{fig:TheovsNum} represent the {\em typical} values for $CV$ and $S$, and that it might be possible to encounter a large deviation from such quantities. Moreover, one might dwell into a parameter regime where such standard deviation bars overlap and hence it is unrealistic to certify the sampler with a single $U_{\rm sub}$ measurement. Luckily, a simple change of input modes implies a change in $U_{\rm sub}$ and hence it is feasible to generate several C-datasets from one circuit.  Figure \ref{fig:VarSkew} shows the outcomes for various such $U_{\rm sub}$ matrices as points, where the x-coordinate indicates the coefficient of variation and the y-coordinate shows the skewness. The colour coded sets of points for different particle types are all separated from each other, showing clearly that they can be distinguished.
As is indicated in Figure  \ref{fig:VarSkew}, upon averaging over all the points in each cloud, one finds values (indicated by the red circles) which are very well estimated by the RMT predictions (indicated by the red triangles), thus providing a strong quantitative tool for such certification. This quality of the certification is further enhanced in large systems by noticing  that the cloud is expected to shrink with the effective number of scattering events inside the array, namely the typical number of crossings between optical paths in Figure \ref{fig:Sketch} (state of the art experiments have $\simeq 20$).

\begin{figure}
	\includegraphics[width=0.49\textwidth]{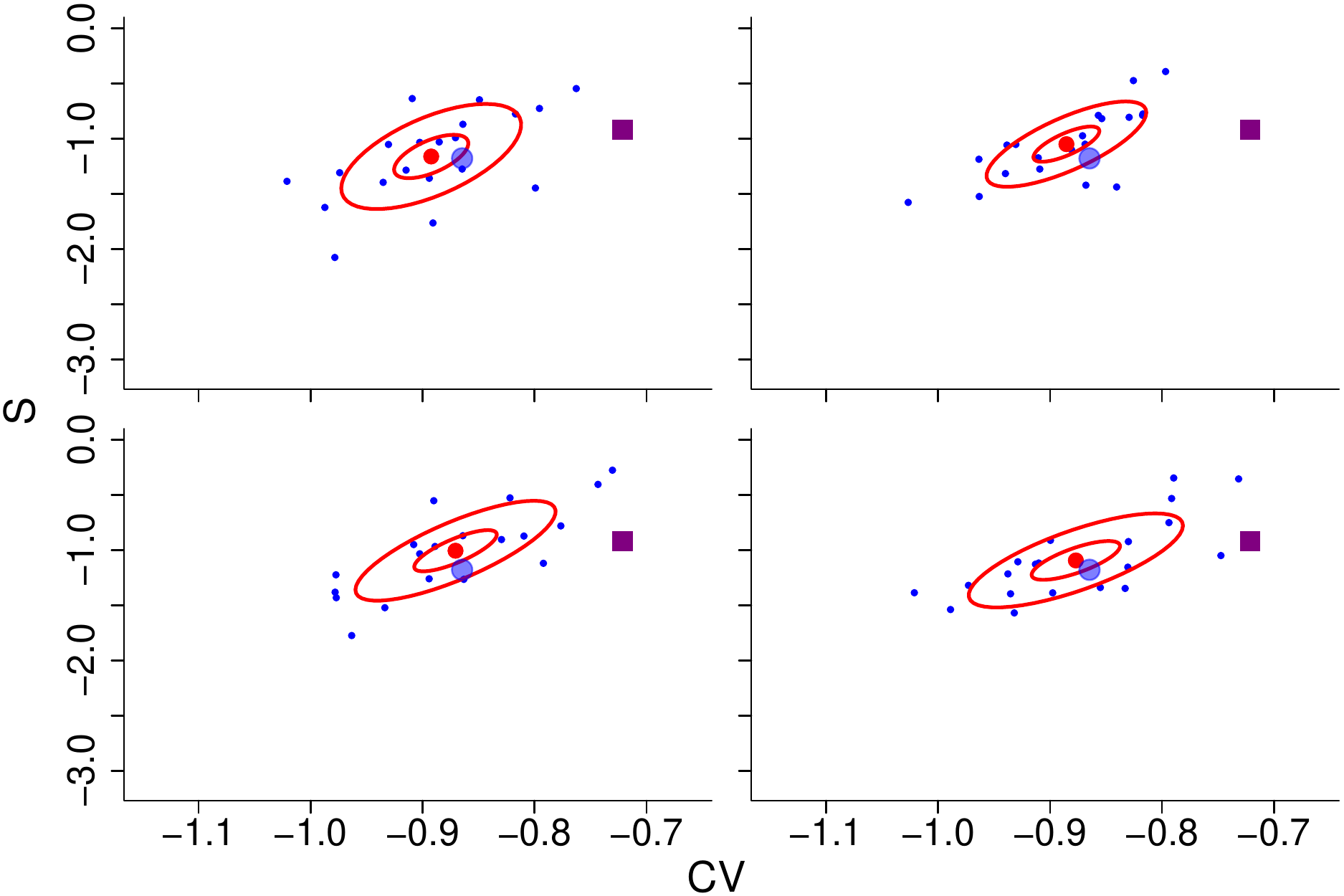}
  \caption{Four scatterplots, each containing 20 randomly sampled $U_{\rm sub}$ matrices for six particles in $m=20$ output modes, from which coefficient of variation $CV$ and the skewness $S$ of the bosonic C-dataset were calculated. The average of the cloud is indicated (red dot) as are the two- and four standard error regions (small and large ellipsoid respectively). The RMT prediction for bosons is shown (large blue dot) to be contained within the ellipses for each of the four samples. The RMT prediction for simulated bosons (purple square) falls well outside the four standard errors.}
   \label{fig:fewmodes}
\end{figure}

In Figure \ref{fig:TheovsNum} we have indicated that bosons, fermions and distinguishable particles can be identified by studying the mean of the C-dataset. Furthermore, Figure \ref{fig:VarSkew} shows that for a rather large number of modes and a large set of sampled $U_{\rm sub}$ matrices, we can classify all species. The method presented, however, does not require such an abundance of modes and samples since we can perform additional statistical analyses on the obtained cluster of data points. We emphasise this in Figure \ref{fig:fewmodes}, where data points for only 20 samples of $U_{\rm sub}$ matrices for $m=20$ are shown. We focus specifically on bosons (indicated by blue points), where for each sample the average is calculated (red dot) and the red ellipses indicate two and four standard errors of the sample mean. The RMT prediction for bosons (a blue circle), with a slight bias, falls within the four standard errors, whereas the RMT prediction for simulated bosons (purple square) is well outside this region. Thus, we can successfully differentiate true bosons from simulated bosons using the RMT-based techniques described here.

\section{Conclusions} As explained above, the suggested methods can be applied in a broad variety of experimental setups, for a wide range of particle numbers $n$ and mode numbers $m$. Of course, smaller $m$ lead to smaller C-datasets, making it difficult to achieve statistical significance. In our numerical studies, however, we successfully distinguish the particle types for as few as 20 modes. A nice advantage of our method as compared to \cite{aaronson_bosonsampling_2013} is that we can certainly treat regimes where $m<n^{5.1}$, we even can explore regimes in which $m \sim {\cal O} (n)$. However, as $n$ and $m$ grow, in Figure \ref{fig:TheovsNum} the curves for true and simulated bosons will approach one another to a distance of the order ${\cal O} (1/n)$. As the limit $1/n \rightarrow 0$ is the semi-classical limit, where mean-field theory is exact, this is as such not surprising. The similarity of the two curves essentially implies that the statistics in the C-dataset is strongly dominated by the quantum statistics of the particles and thus by the bosonic bunching. In this sense, we see a lot of potential in methods such as the one we employed here, to further explore specific signatures of many-body interference, {\em e.g.} by filtering out the bunching contributions to the statistics. \\

In summary, we demonstrated that, by measuring the mode correlators $C_{i,j}=\<\hat{n}_i\hat{n}_j\> -\<\hat{n}_i\>\<\hat{n}_j\> $ for all possible combinations of outgoing modes, one holds the key to certifying BosonSampling. The {\em mean}, the {\em variance}, and the {\em skewness} of such a C-dataset are sufficient to identify the sampled particles as either bosons, fermions or distinguishable particles beyond reasonable doubt. By varying the chosen input channels, and thereby generating multiple such datasets one can efficiently distinguish true bosons from simulated bosons -- which are designed to replicate bunching behaviour -- through the characteristic first moments of their respective distributions by using the RMT predictions presented here, and comparing them to numerically or experimentally obtained results after averaging over the different C-datasets.

What is the consequence thereof for the extended Church-Turing thesis? Clearly, the capacity of any classical computer will be quickly exhausted when confronted with
the task to evaluate a many-body wave function represented by a permanent, as soon as the number of bosonic constituents and modes is large enough. However, much as
in the classical theory of gases or of chaotic (classical or quantum) systems, there are robust statistical quantifiers which indeed can be handled, and which are accessible
in state of the art experiments. In this sense, a ``thermodynamic'' or ``statistical'' interpretation of the extended Church-Turing thesis will prevail.

{\bf Acknowledgements -} M.W. would like to thank the German National Academic Foundation for financial support. M.C.T. acknowledges financial support by the Danish Council for Independent Research. K.M. and A.B. acknowledge partial support through the COST Action MP1006 `Fundamental Problems in Quantum Physics', and by DFG.  J.D.U. and K.R. acknowledge partial support through DFG.

\bibliography{Boson-Sampling-Paper}
\bibliographystyle{apsrev4-1}

\newpage

\onecolumngrid
\appendix
\section{ Correlators }Initially, let us present a short and slightly more technical introduction to the central objects that build up the C-dataset, the two-particle correlators. Given some pure quantum state $\left\lvert\phi\>$, these objects are defined as $C_{ij} = \< \phi \right\rvert \hat{n}_i\hat{n}_j \left\lvert \phi \> -  \< \phi \right\rvert \hat{n}_i\left\lvert \phi \> \< \phi \right\rvert \hat{n}_j \left\lvert \phi \> $, the main goal of studying these object is to gain insight in the structure of states $\left\lvert\phi\>$, which results from the scattering of a many-particle Fock state in a system (one might think of a complicated network) which is described by a single-particle scattering matrix $U$ (which we numerically generate following the algorithm described in \cite{mezzadri_how_2006}). As we initially start from a Fock state for which modes $q_1, \dots, q_n$ are populated by a single particle, we can describe the initial state $\left\lvert \phi_{\rm in} \>$ in terms of creation operators $a^*_q$ (for creation in the $q$th mode) that act on the vacuum state $\left\lvert \Omega \>$, as \begin{equation}
\left\lvert \phi_{\rm in} \> = a^*_{q_1}\dots a^*_{q_n}  \left\lvert\Omega \>.
\end{equation}
Now, by traversing the system, the matrix $U$ acts by connecting an input mode $a^*_q$ to all possible output modes $a^*_i$ \begin{equation}
a^*_q \rightarrow \sum_{i=1}^m U_{q,i}a^*_i
\end{equation}
and thus we obtain that 
\begin{equation}
\left\lvert\phi\> = \sum^m_{i_1, \dots, i_n = 1} U_{q_1,i_1}a^*_{i_1}\dots U_{q_n,i_n}a^*_{i_n} \left\lvert\Omega \>.
\end{equation}
For bosons (B) and fermions (F), an application of the (anti)commutation relations, $[a_i,a^*_j]_{\pm}=\delta_{ij} {\bf 1}$, and a long but straightforward computation leads to expressions for $C_{ij}$:
\begin{align}
& C^B_{ij}= -\sum^n_{k=1} U_{q_k,i}U_{q_k,j}U^*_{q_k,i}U^*_{q_k,j} + \sum^n_{k\neq l=1}U_{q_k,i}U_{q_l,j}U^*_{q_li}U^*_{q_k,j},\\
& C^F_{ij}= -\sum^n_{k=1} U_{q_k,i}U_{q_k,j}U^*_{q_k,i}U^*_{q_k,j} - \sum^n_{k\neq l=1}U_{q_k,i}U_{q_l,j}U^*_{q_l,i}U^*_{q_k,j}.
\end{align}
In the case of distinguishable particles, one can in principle treat the particles in an independent fashion and thus a particle starting in input mode $q$ will be found in output mode $i$ with a probability $p_{q \rightarrow i} = \lvert U_{i,q} \rvert^2$. As the particles are distinguishable, these probabilities are not influenced by the presence of other particles, and via simple probability theory we now find that
\begin{equation}\label{eq:corD}\begin{split}
C^D_{ij} &= \sum^n_{k < l=1} (p_{q_k \rightarrow i}p_{q_l \rightarrow j} + p_{q_k \rightarrow j}p_{q_l \rightarrow i} ) - \sum^n_{k,l=1}p_{q_k \rightarrow i}p_{q_l \rightarrow j}\\
&= -\sum^n_{k=1} U_{q_k,i}U_{q_k,j}U^*_{q_k,i}U^*_{q_k,j}.
\end{split}
\end{equation} 
Finally, simulated bosons behave similarly to distinguishable particles, with the sole exception that the initial state is different and that (uniformly distributed) random phases are included over which one needs to average \cite{tichy_stringent_2014}.  We essentially sample distinguishable particles, which are inserted in the form of a single-particle state that superposes all input modes with the same amplitude, but with random phases, implying a probability $p_{i}= \frac{1}{n} \left\lvert\sum^n_{r=1} e^{i \theta_{q_r}} U_{q_r,i}\right\rvert^2$ to find a particle in output mode $i$. Since every time we consider $n$ such indistinguishable particles, a simple calculation yields \begin{equation}
C^S_{ij} = \mathbb{E}(n(n-1) p_i p_j) -  \mathbb{E}(n p_i) \mathbb{E}(n p_j)
\end{equation}
where $\mathbb{E}(.)$ denotes averaging over the random phases $\theta_r$. Performing the average, we eventually obtain 
\begin{equation}
C^S_{ij} = \left(1-\frac{1}{n}\right) \sum^n_{r \neq s = 1} U_{q_s,i}U_{q_r,j}U^*_{q_r,i}U^*_{q_s,j} - \frac{1}{n} \sum^n_{r,s=1} U_{q_r,i}U_{q_s,j}U^*_{q_r,i}U^*_{q_s,j}.
\end{equation}

\section{Random Matrix Theory } From expressions for the correlators of different particle types, we are able to construct the C-dataset by varying $i$ and $j$ (with $i<j$) to obtain all different mode combinations. In order to do theoretical predictions (or at least up to very good approximation), we use RMT methods. Rather than varying $i$ and $j$, these methods keep the two output modes under consideration fixed and formally average over all possible matrices $U$ in the unitary group with the {\em Haar measure} imposed on it. One might understand this as an analogue to Bohigas-Giannoni-Schmit conjecture \cite{PhysRevLett.52.1} for unitary matrices. The averaging contains one fundamental identity for an $N\times N$ random unitary matrix $U$: \begin{equation}
\mathbb{E}_U(U_{a_1,b_1}\dots U_{a_n,b_n}U^*_{\alpha_1,\beta_1}\dots U^*_{\alpha_n,\beta_n}) = \sum_{\sigma,\pi \in S_n}V_N(\sigma^{-1}\pi)\prod^n_{k=1}\delta(a_k-\alpha_{\sigma(k)})\delta(b_k-\beta_{\pi(k)}),
\end{equation}
where $\mathbb{E}_U(.)$ denotes the average over the unitary group and $V$ are class coefficients also known as Weingarten functions, which are determined recursively, the details of this method can be found in \cite{samuel80, mello90, brouwer_diagrammatic_1996,berkolaiko_combinatorial_2013}. With this formula, we efficiently average long products of coefficients of unitary matrices to compute $\mathbb{E}_U(C_{ij})$, $\mathbb{E}_U(C^2_{ij})$ and $\mathbb{E}_U(C^3_{ij})$ for each particle type. Combining these quantities we can find the coefficient of variation $CV$ and the skewness $S$ for each particle type. We find, with $n$ particles in $m$ modes, for bosons:
\begin{alignat}{1}
&\mathbb{E}_{U}\left({C_B}\right) = \frac{n (-m-n+2)}{m \left(m^2-1\right)},\\
&\mathbb{E}_{U}\left({C_B}^2\right) = \frac{2 n \left(m^2 n+m^2+9 m n-11 m+n^3-2 n^2+5 n-4\right)}{m^2 (m+2) (m+3) \left(m^2-1\right)},\\
&\begin{aligned}\mathbb{E}_{U}\left({C_B}^3\right) = -2n&\left(\frac{m^3 n^2+15 m^3 n+2 m^3+3 m^2 n^3+6 m^2 n^2+213 m^2 n-222 m^2-3 m n^4}{m^2 (m+1) (m+2) (m+3) (m+4) (m+5) \left(m^2-1\right)}\right.\\
&\left.+\frac{45 m n^3+32 m n^2+372 m n-464 m+3 n^5-6 n^4+45 n^3+78 n^2+168 n-288}{m^2 (m+1) (m+2) (m+3) (m+4) (m+5) \left(m^2-1\right)}\right),\end{aligned}
\end{alignat} 
for fermions
\begin{alignat}{1}
&\mathbb{E}_{U}\left({C_F}\right) = \frac{n (n-m)}{m \left(m^2-1\right)},\\
&\mathbb{E}_{U}\left({C_F}^2\right) = \frac{2 n (n+1) (m-n) (m-n+1)}{m^2 (m+2) (m+3) \left(m^2-1\right)},\\
&\mathbb{E}_{U}\left({C_F}^3\right) = -\frac{6 n (n+1) (n+2) (m-n) (m-n+1) (m-n+2)}{m^2 (m+1) (m+2) (m+3) (m+4) (m+5) \left(m^2-1\right)},
\end{alignat} 
for distinguishable particles
\begin{alignat}{1}
&\mathbb{E}_{U}\left({C_D}\right) = -\frac{n}{m (m+1)},\\
&\mathbb{E}_{U}\left({C_D}^2\right) =\frac{n \left(m^2 n+3 m^2+m n-5 m+2 n-2\right)}{m^2 (m+2) (m+3) \left(m^2-1\right)},\\
&\mathbb{E}_{U}\left({C_D}^3\right) = -\frac{n \left(m^2 n^2+9 m^2 n+26 m^2+5 m n^2+21 m n-62 m+12 n^2+60 n-72\right)}{m^2 (m+2) (m+3) (m+4) (m+5) \left(m^2-1\right)},
\end{alignat} 
and finally for the simulated bosons
\begin{alignat}{1}
&\mathbb{E}_{U}\left({C_S}\right) =-\frac{n (m+n-2)}{m \left(m^2-1\right)},\\
&\begin{aligned}\mathbb{E}_{U}\left({C_S}^2\right) = &\frac{4 m n-m-14 n^2+8 n-2}{m^2 (m+2) (m+3) \left(m^2-1\right) n}\\&+\frac{2 m^2 n^3-m^2 n^2+4 m^2 n-m^2+18 m n^3-25 m n^2+2 n^5-4 n^4+10 n^3}{m^2 (m+2) (m+3) \left(m^2-1\right) n},\end{aligned}\\
&\begin{aligned}\mathbb{E}_{U}\left({C_S}^3\right) = &\left(\frac{-2 m^3 n^5-21 m^3 n^4+30 m^3 n^3-41 m^3 n^2-10 m^3 n+8 m^3-6 m^2 n^6-3 m^2 n^5}{(m-1) m^2 (m+1)^2 (m+2) (m+3) (m+4) (m+5) n^2}\right.\\
&\left.+\frac{-285 m^2 n^4+261 m^2 n^3+75 m^2 n^2-66 m^2 n+24 m^2+6 m n^7-90 m n^6-55 m n^5}{(m-1) m^2 (m+1)^2 (m+2) (m+3) (m+4) (m+5) n^2}\right.\\
&\left.+\frac{-360 m n^4+591 m n^3+8 m n^2-128 m n+64 m}{(m-1) m^2 (m+1)^2 (m+2) (m+3) (m+4) (m+5) n^2}\right.\\
&\left.+\frac{-6 n^8+12 n^7-90 n^6-120 n^5-24 n^4+396 n^3-168 n^2-48(n-1)}{(m-1) m^2 (m+1)^2 (m+2) (m+3) (m+4) (m+5) n^2}\right).\end{aligned}
\end{alignat} 
Although these formulas do not appear remarkably elegant due the lack of any form of assumption on $m$ and $n$ (apart from $m>n$), they are necessary to obtain sufficiently accurate results. Once these moments are defined, we can use them to find $NM$, $CV$ and $S$ by the following definitions: \begin{align}
& NM = \frac{\mathbb{E}_U(C) m^2}{n}\\
&CV = \frac{\sqrt{\mathbb{E}_U(C^2)-\mathbb{E}_U(C)^2}}{\mathbb{E}_U(C)},\\
&S = \frac{\mathbb{E}_U(C^3) - 3\mathbb{E}_U(C)\mathbb{E}_U(C^2) + 2\mathbb{E}_U(C)^3}{\left(\mathbb{E}_U(C^2)-\mathbb{E}_U(C)^2\right)^{3/2}}.
\end{align}
With these results, one can now calculate the expected coefficient of variation and the expected skewness for each of the samplers we described, with an arbitrary number of modes and particles.

\end{document}